# Fully phase-stabilized quantum cascade laser frequency comb

Luigi Consolino,[1*] Malik Nafa,[1] Francesco Cappelli,[1] Katia Garrasi,[2] Francesco P. Mezzapesa,[2] Lianhe Li,[3] A. Giles Davies,[3] Edmund H. Linfield,[3] Miriam S. Vitiello,[2] Paolo De Natale[1] and Saverio Bartalini[1,4]

[1]CNR-Istituto Nazionale di Ottica and LENS, Via N. Carrara 1, 50019 Sesto Fiorentino (FI), Italy

[2]NEST, CNR - Istituto Nanoscienze and Scuola Normale Superiore, Piazza S. Silvestro 12, 56127, Pisa, Italy

[3]School of Electronic and Electrical Engineering, University of Leeds, Leeds LS2 9JT, UK

[4]ppqSense Srl, Via Gattinella 20, 50013 Campi Bisenzio FI, Italy

[*]luigi.consolino@ino.it

**Optical frequency comb synthesizers (FCs)[1] are laser sources covering a broad spectral range with a number of discrete, equally spaced and highly coherent frequency components, fully controlled through only two parameters: the frequency separation between adjacent modes and the carrier offset frequency. Providing a phase-coherent link between the optical and the microwave/radio-frequency regions,[2] FCs have become groundbreaking tools for precision measurements.[3,4]**

**Despite these inherent advantages, developing miniaturized comb sources across the whole infrared (IR), with an independent and simultaneous control of the two comb degrees of freedom at a metrological level, has not been possible, so far. Recently, promising results have been obtained with compact sources, namely diode-laser-pumped microresonators[5,6] and quantum cascade lasers (QCL-combs).[7,8] While both these sources rely on four-wave mixing (FWM) to generate comb frequency patterns, QCL-combs benefit from a mm-scale miniaturized footprint, combined with an *ad-hoc* tailoring of the spectral emission in the 3-250 μm range, by quantum engineering.[9]**

**Here, we demonstrate full stabilization and control of the two key parameters of a QCL-comb against the primary frequency standard. Our technique, here applied to a far-IR emitter and open ended to other spectral windows, enables Hz-level narrowing of the individual comb modes, and metrological-grade tuning of their individual frequencies, which are simultaneously measured with an accuracy of $2\times10^{-12}$, limited by the frequency reference used. These fully-controlled, frequency-scalable, ultra-compact comb emitters promise to pervade an increasing number of mid- and far-IR applications, including quantum technologies, due to the quantum nature of the gain media.[10]**

Thorough characterizations of QCL-combs, recently performed both in the mid-IR[11,12] and in the far-IR,[12,13] have demonstrated that FWM is a strong mode-locking mechanism, providing tight phase relation among the modes simultaneously emitted by the devices. However, while QCL-based FCs have been already used in a variety of sensing setups, mostly related to dual-comb spectroscopy,[14,15] their relevance as metrological-grade, phase-stabilized sources has not been proven, yet. In fact, simultaneous stabilization of the two key comb degrees of freedom requires two suitable and independent actuators. Although reliable procedures for stabilizing the QCL-comb modes spacing[7,16] or the frequency of a single mode[17] have been reported, independent tuning of carrier offset and spacing, together with an effective overall phase stabilization has represented, for a long time, a non-trivial breakthrough.

A detailed description of the QCL used in the present work can be found in Methods section M.1. The bias-dependent emission bandwidth displays a maximum frequency coverage of 1.3 THz,[18] and is centered at 2.9 THz. The mode spacing, defined by the cavity length, can be electrically extracted through the measurement of the intermode beat-note (IBN - $f_{IBN}$ ~17.45 GHz).

The experimental setup used for stabilization and characterization of the QCL-comb emission is based on a multi-heterodyne detection scheme between the QCL-comb and a fully-stabilized free-standing optically-rectified THz frequency comb[19,20] (OR-comb) (see Methods section M.2). The frequency mixing results in a down-converted radio-frequency comb (RF-comb), that carries full information on the QCL-comb emission.[12]

Fig. 1a shows the RF-comb spectrum of the free-running QCL-comb (blue trace) acquired by a real-time spectrum analyzer. The beat-notes (BNs) pattern distinctly retraces the QCL-comb Fourier transform infrared (FTIR) spectrum measured in air (Methods section M.1). In these conditions, the emission linewidth of the OR-comb modes does not contribute to the BNs width, which shows a linear increase (about 14.4 kHz per mode) with the QCL-comb order $N$. This increment is close to the IBN width for a comparable acquisition time, confirming that the observed trend can be ascribed to frequency spacing fluctuations. In this scenario, a full active stabilization of the comb emission frequencies can be performed by means of two independent actuators.

On one hand, the mode spacing of the QCL-comb can be stabilized through injection locking (fig. 1b), by feeding to the QCL a radio-frequency signal ($f_{RF}$) approaching the QCL cavity round-trip frequency (see section M.2). When the injection locking is active, the QCL-comb frequency spacing fluctuations are drastically reduced (fig. 1b), while only the mode-independent offset-frequency fluctuations survive, as confirmed by the roughly constant width (120 kHz) of the BN signals (red dots).

On the other hand, any mode of the RF-comb can be singly stabilized to a reference oscillator ($f_{LO}$), using a phase-lock loop (PL-loop) acting on the QCL driving current (fig.1c) (see Methods M.2). In this case, in absence of the QCL-comb mode spacing stabilization, the BN linewidths linearly increase with the order $N$ of the comb tooth, with a slope of about 66.5 kHz/mode, much larger than that retrieved in the free running conditions (Fig. 1a), since the offset fluctuations of the stabilized mode propagate to the adjacent modes as spacing fluctuations. Consequently, a visible broadening is observed for the BNs distant from the stabilized mode.

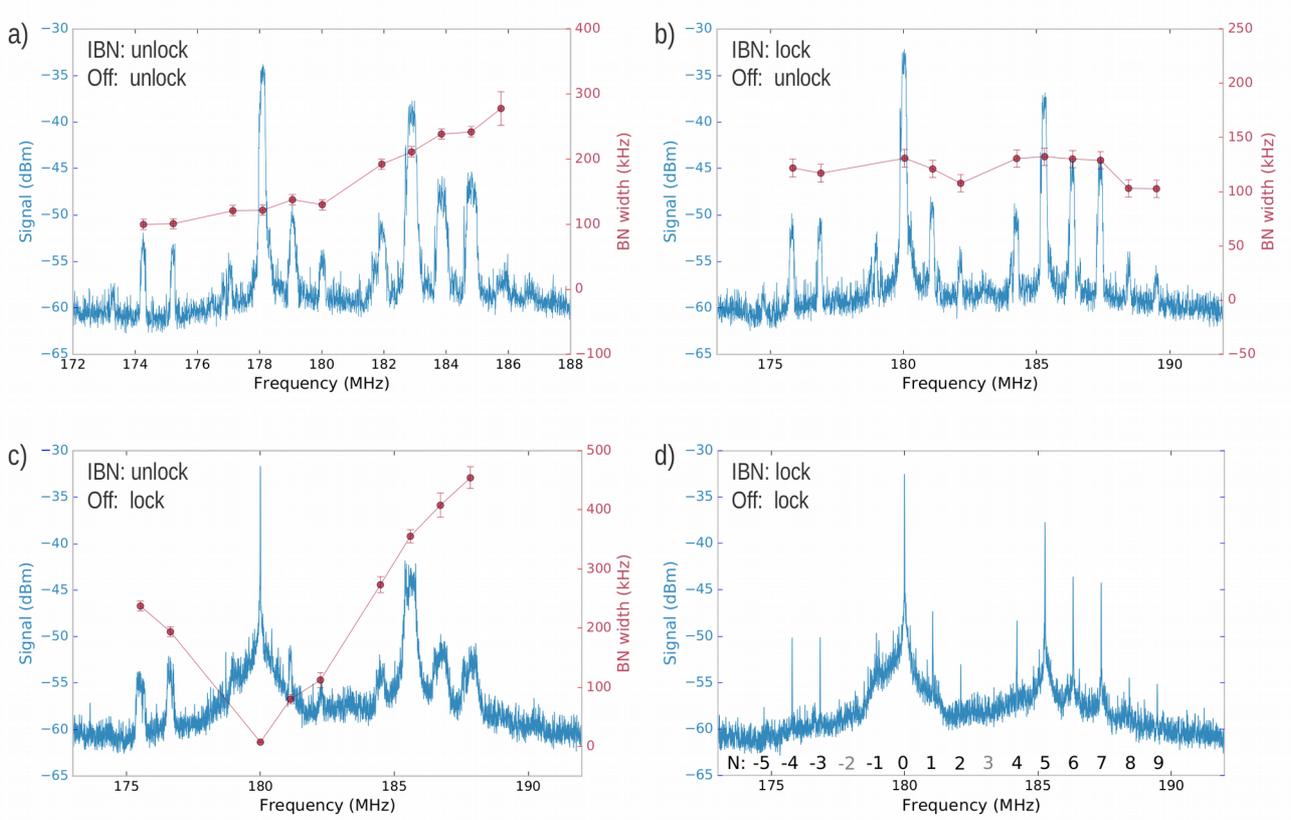

Fig. 1: **Radio frequency spectra.** Radio-frequency (RF) comb spectra (blue traces) and related width of each beat-note (red dots). In all the acquisitions the optically-rectified THz frequency comb is fully stabilized, while the QCL-comb is: (a) free-running; (b) stabilized mode spacing and free-running offset frequency; (c) free-running spacing and stabilized offset; (d) fully stabilized. The order $N$ of each comb mode is labeled on the last graph.

Finally, the simultaneous implementation of both these processes results in a full stabilization of the QCL-comb (shown in fig. 1d), as confirmed by the linewidth of the acquired BN signals, that are all resolution bandwidth (RBW) limited. It is worth noticing that both the stabilization signals are sent to the QCL-comb as electrical signals, acting on its voltage/current, over two well separated frequency ranges: the RF range for the 17.45 GHz signal that locks the mode spacing, and the DC/low frequency range for the PL-loop, with a 400 kHz bandwidth. Actually, this configuration only requires a single-frequency metrological-grade THz signal,[21–23] falling into the QCL-comb emission spectrum. In this work, we have used one mode of the OR-comb, exploiting its complete emission for an exhaustive characterization of the QCL-comb.

The QCL-comb mode frequencies can be simultaneously measured with high accuracy by acquiring RF spectra with a small RBW (0.5 Hz), limited by the spectrum analyzer maximum acquisition time of 2 seconds (fig. 2a). The long acquisition time allows lowering the noise floor, thus enabling the detection of the 19 BNs corresponding to all the emitted modes. Their RF frequencies coincide, within the RBW (fig. 2a, insets), with the ones extrapolated from the reference signals frequencies (see Methods section M.2). As a consequence, the emission linewidth of each QCL-comb mode is narrowed down to the linewidth of the OR-comb modes. This latter is ~2 Hz in 1 s, limited by the stability of the GPS-Rb-clock frequency reference used. The frequency reference accuracy of $2\times10^{-12}$ also limits the measurement of the absolute QCL-comb THz frequencies, to a 6 Hz-level. However, for the most demanding applications, the GPS chain can be replaced with a fiber-delivered optical frequency standard, already available in our laboratories,[24] providing a stability of $1\times10^{-14}$ in 1 s, and an accuracy of $2\times10^{-16}$.[25]

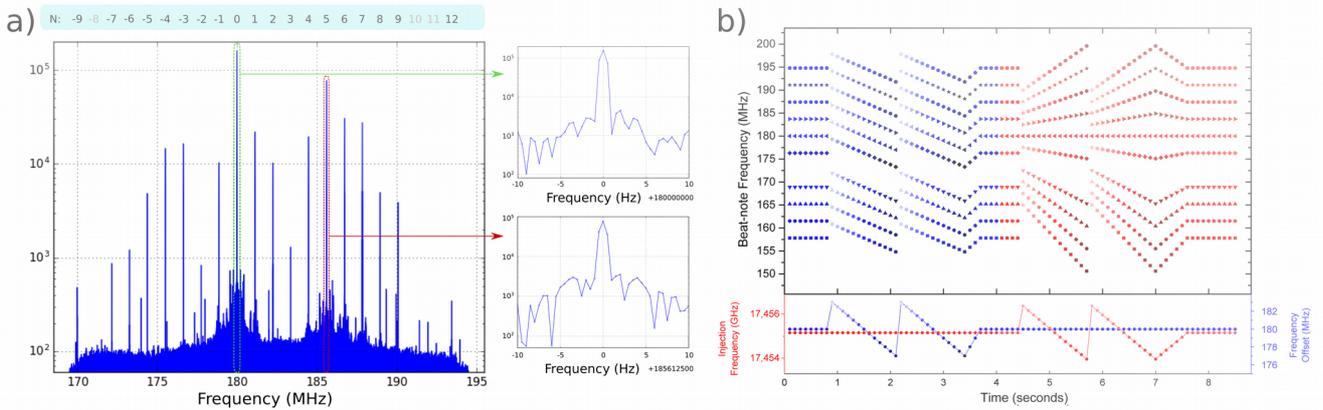

Fig. 2: **Down-converted comb spectra and frequency tuning.** (a) RF-comb spectrum measured with 0.5 Hz RBW, limited by the acquisition time of 2 seconds. Insets: zoom on the RBW-limited peaks. (b) Fine tuning of the QCL-comb frequencies. First, the $f_{LO}$ frequency of the PL-loop that stabilizes the comb-offset is tuned, in steps of 500 kHz, between 183 and 177 MHz. Then, the frequency of the injecting RF-signal ($f_{RF}$) is tuned by 2.4 MHz, in steps of 200 kHz. Simultaneously, the RF-comb spectrum is acquired and the frequencies of individual modes are measured with a 100 Hz RBW.

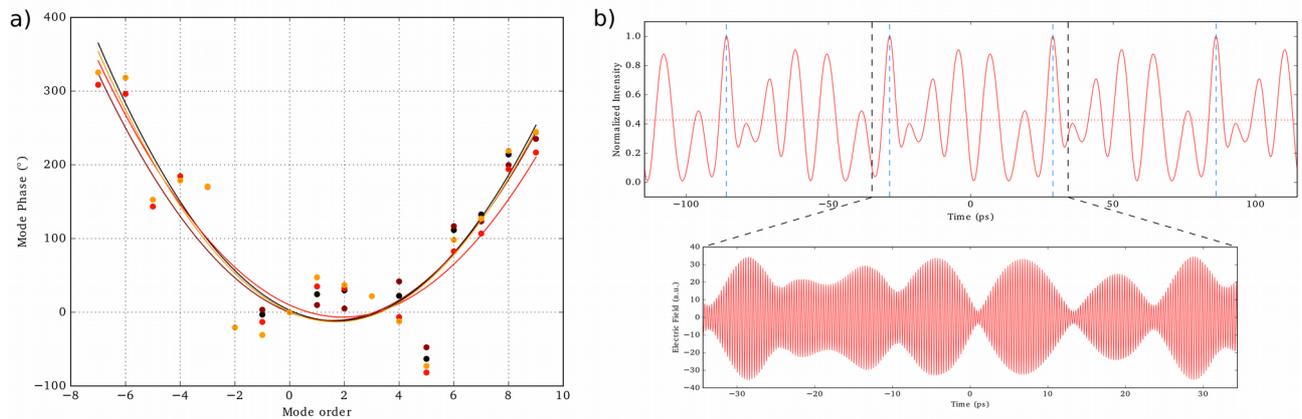

Fig. 3: **Measured phases.** (a) Measured phases of each emitted mode. Different colors represent sets of measurements collected during different days and under different operating conditions. (b) QCL output intensity profile and electric field, reconstructed through one set of phases shown in panel (a).

By varying $f_{RF}$ and $f_{LO}$, a fine tuning of the corresponding QCL-comb parameters (spacing and offset, respectively) is obtained, as confirmed by the retrieved shifts of the RF-comb frequencies (fig. 2b). Therefore, the double-locking procedure ensures the tuning of the QCL-comb along two independent axes of its two-dimensional parameters space, that is a key requisite for full practical exploitation of this metrological-grade source. Actually, although the two actuators are not perfectly "orthogonal", since the driving current affects both the spacing and offset frequencies, injection locking acts on the mode spacing only, allowing an independent tuning of the two parameters.

The multi-heterodyne FFT analysis also enables retrieving the complete set of the modal phases, allowing to investigate their short-term noise and their long-term stability and reproducibility. To this purpose, several 2-seconds-long acquisitions of the multi-heterodyne signal have been analyzed with the technique described in ref.[12]. In order to assess the long-term reproducibility of the phases, different measurements were performed over different days, power cycling the QCL, changing the injection locking frequency and the device operating parameters (i.e. temperature and driving current). The device shows a quite stable phase pattern (fig. 3a), with a phase reproducibility for individual modes always better than 50°. The retrieved phase relations show a parabolic trend, giving a 6.8(0.5) ps$^2$/rad group delay dispersion (GDD). This value is close to that measured for a QCL-comb operating under comparable band alignment conditions, and without dispersion compensation geometry.[12] The retrieved phases and amplitudes of the QCL modes can also be used to reconstruct the QCL electric field and intensity emission profile, as shown in fig. 3b. These reconstructions confirm that, while the QCL intensity is deeply modulated in time, it is far different from a short-pulse emission, since light is emitted during more than the 40% of the round-trip time.

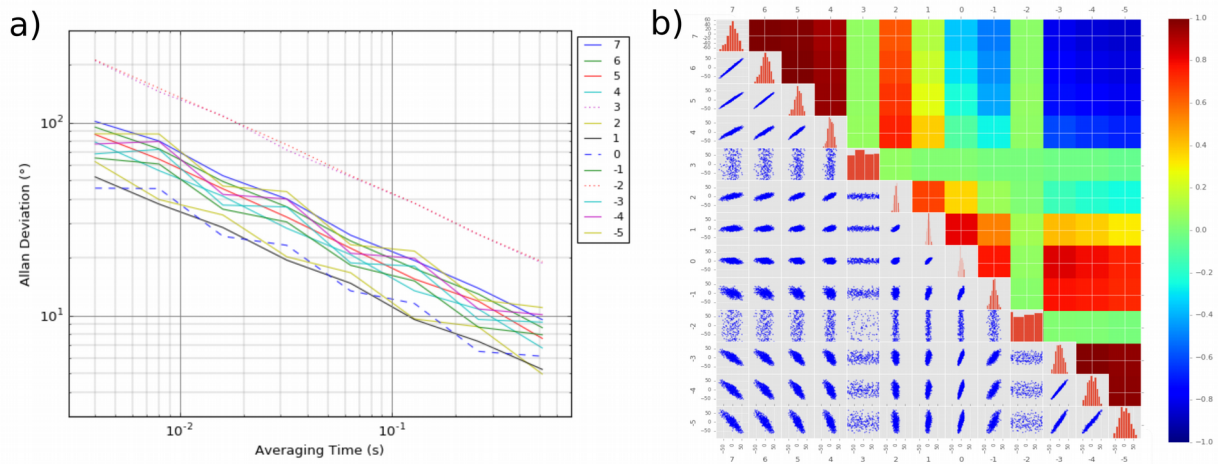

Fig. 4: **Analysis of the phase noise** (a) Allan variance analysis on the phase of each comb mode during a 2-seconds-long acquisition. The characteristic trend confirms that the measurements are affected by statistical noise. (b) Correlation analysis on the phases noise fluctuations. The correlation among the most intense modal phases confirms that the statistical noise detected is due to residual instabilities of the QCL-comb spacing.

The short-term stability of the phases has been evaluated through calculation of the Allan deviations over a 2-seconds-long measurement for the modes labeled as -5 ... 7 (fig.1d). The unveiled $1/\tau^{1/2}$ trends (fig. 4a) and the absence of visible sharp features confirm that the phase measurements are white-noise limited. A larger noise level is found for modes -2 and 3, that have the weakest S/N under the operating bias/temperature conditions (see fig. 2a). Except from these two modes, the Allan deviations of the phases at ~1 s are all within 10 degrees, which is a remarkable result if compared with the state-of-the-art in other spectral regions.[26] Moreover, the lowest noise is measured on the phase-locked mode 0 and its neighbors, and progressively increases with the order N. This suggests that the dominant noise source is residual noise from the spacing locking mechanism (i.e. injection locking), that coherently propagates along the comb. Such effect can be evidenced by calculating the correlation matrix among the modal phases (fig. 4b). For modes -2 and 3, no correlation with the other modes appears, confirming that their fluctuations are limited by measurement statistical noise. At the same time, strong correlations and anti-correlations between modes which are, respectively, adjacent and opposite with respect to mode 0 can be seen. This confirms the high level of coherence of the QCL-comb, and that a residual collective phase noise mainly affects comb spacing, rather than carrier offset frequency, suggesting that the injection locking mechanism should be improved.

In conclusion, the reported results show that QCL-combs are well suited for the most demanding applications, achieving a Hz-level stabilization of each individual emitted mode, as well as a complete and independent control of the two key comb parameters. The retrieved phase relation, that enables reconstruction of the electric field and of the intensity profile, is impressively reproducible over different days, power cycles and operating conditions. The high degree of coherence induced by the FWM mode-locking mechanism is confirmed by a correlation analysis. A residual collective phase noise is observed, due to a non-perfect injection locking mechanism. In the next future, such a double-locking and phase characterization setup will be applied to dispersion-compensated devices, where comb operation at higher operating currents will ensure a broader spectral coverage. Novel comb-based setups will benefit from such fully phase-stabilized QCL-combs, addressing the most demanding metrological and sensing applications with miniaturized devices emitting in key spectral windows difficult to access, so far. Moreover, harnessing QCL nonlinearity, exploiting also novel phenomena, like $\chi^2$ cascaded frequency generation,[27] and making use of quantum simulation schemes in view of a fully-quantum design, QCL-combs with entangled modes and noise squeezed emission could become available, providing a solid-state miniaturized platform with a huge impact on a wide range of quantum technologies.

## Methods

**M.1: Design and fabrication of the QCL comb.**

The heterogeneous GaAs/AlGaAs heterostructure comprises three active modules, grown on a semi-insulating GaAs substrate by molecular beam epitaxy, exploiting alternating photon- and longitudinal optical (LO) phonon-assisted transitions between inter-miniband.[18] The number of periods, the order of the active region modules and the average doping were carefully arranged to give a flat gain and uniform power output across the whole spectrum. The gain medium of each modules is respectively centered at 3.5 THz, 3.0 THz and 2.5 THz . The average doping was set to $3\times10^{16}$ cm$^{-3}$.[28] Laser bars were fabricated on a standard metal–metal processing that relies on Au-Au thermo-compression wafer bonding of the 17-µm-thick active region onto a highly doped GaAs substrate. Laser bars were defined using dry-etching, leading to almost vertical sidewalls narrow laser cavities without any unpumped regions, to favour stable continuous wave (CW) operation. A Cr/Au (8 µm/150 µm) top contact was then deposited on the top surface of the laser bars, leaving thin (5 µm) symmetrical side regions uncovered, to suppress undesired high order lateral modes. Thin (5-nm-thick) lateral nickel side absorbers were then deposited over the uncovered region using a combination of optical lithography and thermal evaporation. Laser bars 60–85 µm wide and 2 mm long were finally cleaved and mounted on a copper bar on the cold finger of a helium cryostat.

Fig. M1a shows the current-voltage characteristics of the THz QCL-comb used in this work. The intermode beat-note retrieved at 320 mA, where the QCL shows a single narrow signal, is plotted in Fig. M1b, while the corresponding in air Fourier transform infrared spectrum is graphed in panel M1c, for reference.

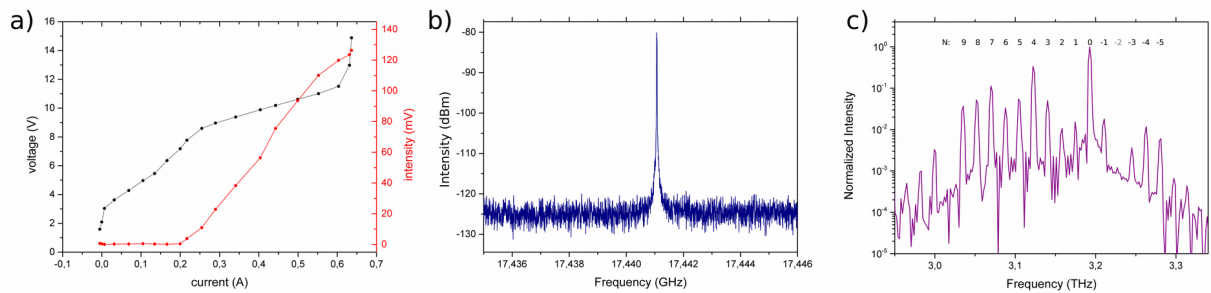

Fig. M1: **Electrical and optical characterization of the QCL-comb.** (a) current-voltage characteristics of the THz QCL-comb. (b) intermode beat-note (IBN) spectrum measured at a heat sink temperature of 25 K and with QCL operating in continuous wave with a current of 0.32 A. The IBN is 20 kHz wide at 20ms integration time. (c) Fourier transform infrared spectrum measured in rapid scan mode, in air, with a resolution of 0.125 cm$^{-1}$. The order N of each mode is labeled on the graph.

**M.2: Dual-comb**

The experimental setup used for full stabilization and characterization of the QCL-comb is shown in Fig. M2a. It is based on multi-heterodyne detection, a technique employed in dual-comb spectroscopy experiments.[14] An amplified mode-locked Erbium doped fiber fs laser (Menlo System, model FC1500) is optically rectified in a single-mode Lithium Niobate waveguide,[22] providing a zero-offset free-standing THz frequency comb (OR-comb)[19]. The repetition rate ($f_{rep}$) of the pump laser can be tuned around 250 MHz, being the same for the generated THz OR-comb, while the average power of 350 mW at 1.55 μm generates an overall THz average power of about 14 μW. The QCL is mounted on the cold finger of a liquid Helium cryostat at a fixed heat sink temperature of 25 K, and it is driven in CW mode by an ultra-low noise current driver (ppqSense, QubeCL-P05) around 0.32 A. The OR-comb and the QCL-comb beams are overlapped by means of a wire grid polarizer (WGP) and generate the multi-heterodyne signal on a fast non-linear detector, i.e. a hot electron bolometer (HEB).

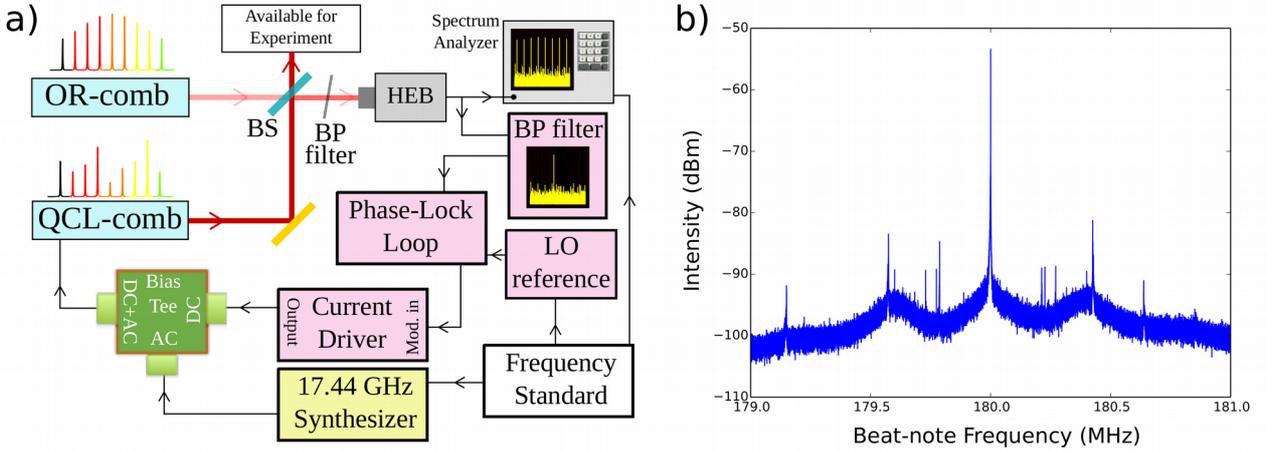

Fig. M2. **Experimental set-up**. a) Schematics of the experimental setup employed for full stabilization, control and characterization of the QCL-comb. The beams of the OR-comb and QCL-comb are superimposed by means of a beam splitter (BS) and then mixed on a fast detector (HEB: hot-electron bolometer). The HEB signal is acquired on a spectrum analyzer (Tektronics RSA5106A) while the two stabilization loops act on the QCL bias. The yellow background object represents the RF synthesizer used for injection locking, while the pink background objects are related to offset stabilization through phase-lock loop (BP filter: band-pass filter, LO reference: local oscillator reference). All the oscillators and the spectrum analyzer are frequency-referenced to the primary frequency standard. b) Phase-locked beat-note signal acquired with a 2 MHz span and 10 Hz RBW. The two sidebands indicate a phase-lock electronic bandwidth of 400 kHz.

The repetition rate $f_{rep}$ of the OR-comb is tuned to be close to an integer submultiple of the frequency spacing $f_s$ of the QCL modes, with a small detuning $f_d$:

$$f_s = k \cdot f_{rep} + f_d \qquad \text{eq.1.}$$

In this configuration, the QCL modes at THz frequencies are down-converted to RF beat-note (BN) signals. In fact, considering the electrical field of the $N^{th}$ mode of the QCL-comb:

$$E_{Q,N} = B_N e^{i[2\pi(f_{off}+Nf_s)t+\varphi_N]} \qquad \text{eq.2}$$

heterodyning it with the M$^{th}$ mode of the OR-comb:

$$E_{O,M} = A_M e^{i[2\pi(Mf_{rep})t+\varphi_M]} \qquad \text{eq. 3,}$$

being $M = k \cdot N$, the down-converted signal will carry the term:

$$A_M B_N e^{i[2\pi(f_{off}+Nf_s-Mf_{rep})t+(\varphi_N-\varphi_M)]} = A_M B_N e^{i[2\pi(f_{off}+Nf_d)t+(\varphi_N-\varphi_M)]} \qquad \text{eq. 4.}$$

In the above equations, $f_{off}$ is the offset frequency of the QCL-comb (the OR-comb is offset-free), while $\varphi_N$ and $\varphi_M$ are the Fourier phases of the QCL-comb mode and of the OR-comb mode, respectively.

If $f_d$ is chosen to be in the range of few MHz, the resulting BNs are RF signals, equally spaced by $f_d$, that can be simultaneously analyzed by a spectrum analyzer. The signal in eq. 4 clearly bears information on both the amplitudes and phases of the two parent modes. In particular, the OR-comb phases are constant in time and their phase relation is linear, as for any mode-locked pulsed FC source,[12] while their amplitude is expected to be almost constant along the QCL-comb emitted spectrum.[19] Moreover, the OR-comb is naturally a zero-offset FC, whose modes can be fully stabilized at a metrological level by phase-locking its repetition rate to a 10 MHz quartz-clock disciplined by a Rb-GPS (global positioning system) clock with a stability of 6×10$^{-13}$ in 1 s, and absolute accuracy of 2×10$^{-12}$. On the other hand, the free-running QCL-comb emission frequencies will oscillate due to both offset and spacing fluctuations, preventing an accurate phase/frequency retrieval from the BN signals, especially for long observation times. The electronic setups sketched in fig. M2a are used to remove these fluctuations, fully stabilizing the QCL-comb emission.

A frequency synthesizer, oscillating at a frequency close to the QCL cavity round trip time (around 17.45 GHz) is used for injection locking the QCL-comb spacing, while a phase-locked loop (PL-loop) circuit, on board the current driver, is used to stabilize the frequency of one down-converted mode against a local oscillator frequency around 180 MHz. The phase-lock loop has a 400 kHz electronic bandwidth and is capable of effectively squeezing the power of the QCL-comb mode down to the OR-comb mode width (see fig. M2b). The two signals are coupled to the QCL by a cryogenic bias-tee (Marki Microwave, mod. BT-0024SMG) placed on chip, very close to the QCL device. The synthesizer, the PL-loop local oscillator and the spectrum analyzer used for the acquisitions are tightly ruled by the GPS-Rb-quartz chain.

Moreover, since all the parameters governing the QCL-comb down conversion are referenced to the primary frequency standard, the frequency $f_i$ of the i$^{th}$ RF-comb mode can be calculated with this same precision, simply considering that:

$$f_i = f_{LO} + i \cdot (k f_{rep} - f_{RF}) \qquad \text{eq. 5,}$$

and these values can be compared to the ones experimentally retrieved with a 500 mHz accuracy, as reported in the main text.

This accuracy level allows also to simultaneously retrieve all the absolute THz frequencies of the QCL-comb modes with an accuracy limit set by the GPS-Rb-clock chain. In fact, if the RF beat-note at frequency $f_i$ is originated from the $N^{th}$ mode of the QCL comb, beating with the $M^{th}$ mode of the OR-comb, one can write the relation:

$$f_{QCL,N} = M \cdot f_{rep} \pm f_i \qquad \text{eq. 6.}$$

By tuning $f_{rep}$ and leaving all the other parameters unaltered, it is possible to retrieve the order M (around 12000), as thoroughly explained in Ref. [29]. Finally, while the accuracy on $f_i$ is 500 mHz, the phase-locked $f_{rep}$ follows the GPS-Rb-quartz chain accuracy, that provides the limiting uncertainty factor in the $f_{QCL,N}$ measurement of about 6 Hz.

**Acknowledgements**

The authors acknowledge financial support from the EC Project 665158 (ULTRAQCL), the ERC Project 681379 (SPRINT), European ESFRI Roadmap (Extreme Light Infrastructure – ELI), EC–H2020 Laserlab-Europe grant agreement 654148, Ministero dell'Istruzione, dell'Università e della Ricerca (Project PRIN-2015KEZNYM "NEMO – Nonlinear dynamics of optical frequency combs").

**Authors contributions**

L.C. and S.B. conceived the experiment, L.C. performed the measurements, L.C., M.N., F.C. and S.B analysed the data, L.C. wrote the manuscript, M.N., F.C., M.S.V., P.D.N. and S.B. contributed to manuscript revision, K.G., F.P.M. and M.S.V. fabricated the QCL device, L.L., A.G.D. and E.H.L. grew the QCL heterostructure, L.C., M.N., F.C., P.D.N. and S.B. discussed the results, M.S.V. supervised the QCL fabrication work, all the other work was done under the joint supervision of P.D.N. and S.B..

**Additional information**

**Competing financial interests:** The authors declare no competing financial interests.

**Data availability:** The data supporting the findings of this study are available from the corresponding author upon reasonable request.

**References**

1. Hänsch, T. W. Nobel lecture: Passion for precision. *Rev. Mod. Phys.* **78,** 1297–1309 (2006).

2. Diddams, S. A. *et al.* Direct link between microwave and optical frequencies with a 300 THz femtosecond laser comb. *Phys. Rev. Lett.* **84,** 5102–5105 (2000).

3. Udem, T., Holzwarth, R. & Hänsch, T. W. Optical frequency metrology. *Nature* **416,** 233–237 (2002).

4. Holzwarth, R. *et al.* Optical frequency synthesizer for precision spectroscopy. *Phys. Rev. Lett.* **85,** 2264–2267 (2000).

5. Savchenkov, A. A. *et al.* Generation of Kerr combs centered at 4.5 μm in crystalline microresonators pumped with quantum-cascade lasers. *Opt. Lett.* **40,** 3468 (2015).

6. Luke, K., Okawachi, Y., Lamont, M. R. E., Gaeta, A. L. & Lipson, M. Broadband mid-infrared frequency comb generation in a $Si_3N_4$ microresonator. *Opt. Lett.* **40,** 4823 (2015).

7. Hugi, A., Villares, G., Blaser, S., Liu, H. C. & Faist, J. Mid-infrared frequency comb based on a quantum cascade laser. *Nature* **492,** 229–233 (2012).

8. Burghoff, D. *et al.* Terahertz laser frequency combs. *Nat. Photonics* **8,** 462–467 (2014).

9. Li, H. *et al.* Dynamics of ultra-broadband terahertz quantum cascade lasers for comb operation. *Opt. Express* **23,** 33270 (2015).

10. Vitiello, M. S. *et al.* Quantum-limited frequency fluctuations in a terahertz laser. *Nat. Photonics* **6,** 525–528 (2012).

11. Singleton, M., Jouy, P., Beck, M. & Faist, J. Evidence of linear chirp in mid-infrared quantum cascade lasers. *Optica* **5,** 948 (2018).

12. Cappelli, F. *et al.* Retrieval of phase relation and emission profile of quantum cascade laser frequency combs. Submitted (2018).

13. Burghoff, D. *et al.* Evaluating the coherence and time-domain profile of quantum cascade laser frequency combs. *Opt. Express* **23,** 1190 (2015).

14. Coddington, I., Newbury, N. & Swann, W. Dual-comb spectroscopy. *Optica* **3,** 414 (2016).

15. Villares, G., Hugi, A., Blaser, S. & Faist, J. Dual-comb spectroscopy based on quantum-cascade-laser frequency combs. *Nat. Commun.* **5,** 5192 (2014).

16. Hillbrand, J., Maxwell Andrews, A., Detz, H., Strasser, G. & Schwarz, B. Coherent injection locking of quantum cascade laser frequency combs. *Nat. Photonics* (2018). doi:10.1038/s41566-018-0320-3

17. Cappelli, F. *et al.* Frequency stability characterization of a quantum cascade laser frequency comb. *Laser Photonics Rev.* **10,** 623–630 (2016).

18. Garrasi, K. *et al.* High dynamic range, heterogeneous, terahertz quantum cascade lasers featuring thermally-tunable frequency comb operation over a broad current range. *ACS Photonics* Submitted, (2018).

19. Consolino, L. *et al.* Phase-locking to a free-space terahertz comb for metrological-grade terahertz lasers. *Nat. Commun.* **3,** 1040 (2012).

20. Bartalini, S. *et al.* Frequency-comb-assisted terahertz quantum cascade laser spectroscopy. *Phys. Rev. X* **4,** 21006 (2014).

21. Consolino, L., Bartalini, S. & De Natale, P. Terahertz Frequency Metrology for Spectroscopic Applications: a Review. *J. Infrared, Millimeter, Terahertz Waves* **38,** 1289–1315 (2017).

22. De Regis, M., Consolino, L., Bartalini, S. & De Natale, P. Waveguided Approach for Difference Frequency Generation of Broadly-Tunable Continuous-Wave Terahertz Radiation. *Appl. Sci.* **8,** 2374 (2018).

23. De Regis, M. *et al.* Room-Temperature Continuous-Wave Frequency-Referenced Spectrometer up to 7.5 THz. *Phys. Rev. Appl.* **10,** 064041 (2018).

24. Insero, G. *et al.* Measuring molecular frequencies in the 1-10 µm range at 11-digits accuracy. *Sci. Rep.* **7,** 12780 (2017).

25. Calonico, D. *et al.* High-accuracy coherent optical frequency transfer over a doubled 642-km fiber link. *Appl. Phys. B Lasers Opt.* **117,** 979–986 (2014).

26. Chen, Z., Yan, M., Hänsch, T. W. & Picqué, N. A phase-stable dual-comb interferometer. *Nat. Commun.* **9,** 3035 (2018).

27. Mosca, S. *et al.* Modulation Instability Induced Frequency Comb Generation in a Continuously Pumped Optical Parametric Oscillator. *Phys. Rev. Lett.* **121,** 093903 (2018).

28. Li, L. H. *et al.* Broadband heterogeneous terahertz frequency quantum cascade laser. *Electron. Lett.* **54,** 1229–1231 (2018).

29. Consolino, L. *et al.* Spectral purity and tunability of terahertz quantum cascade laser sources based on intracavity difference-frequency generation. *Sci. Adv.* **3,** e1603317 (2017).